\documentclass{article}

\usepackage{algorithm}
\usepackage{algorithmic}

\usepackage{mathtools,amssymb,amsmath}
\usepackage{color,bm,xcolor}
\usepackage{multirow}
\usepackage{bm}
\usepackage{indentfirst}
\usepackage{natbib}
\usepackage{mathrsfs}
\usepackage{caption} 
\usepackage{float}
\usepackage{authblk}
\usepackage{graphicx}
\usepackage{subcaption}
\usepackage{setspace}

\newcommand{\hei}[1]{\color{black}{#1} \color{black} }

\definecolor{brown}{rgb}{0.8, 0.33, 0.1}

\newcommand{\hong}[1]{\color{black}{#1} \color{black}  }
\newcommand{\yy}{\hong \it}
\newcommand{\jj}{\hei \rm}

\newcommand{\om}[1]{\Omega_{ij}^{(#1)}}
\newcommand{\cA}{{\cal A}}

\newcommand*\samethanks[1][\value{footnote}]{\footnotemark[#1]}
{
	\title{\bf The    Ci3+3 Design for Dual-Agent Combination    Dose-Finding Clinical Trials   }

	\author{
		Shijie Yuan\thanks{Laiya Consulting Inc., Chicago, USA} \ , Tianjian Zhou\thanks{Department of Statistics, Colorado State University, Fort Collins, USA} \ , Yawen Lin\samethanks[1] \ , and Yuan Ji\thanks{Department of Public Health Sciences, The University of Chicago, Chicago, USA}
	}
	\date{\today}
}

\begin{document}
\begin{titlepage}
	\maketitle
	\begin{abstract}
		We propose a rule-based statistical design for combination dose-finding trials with two agents. The Ci3+3 design is an extension of the i3+3 design with simple 
		up-and-down  decision rules  comparing the observed toxicity rates and equivalence intervals that define the maximum tolerated dose combination. Ci3+3 consists of two stages to allow fast and efficient exploration of the dose-combination space. Statistical inference is restricted to a beta-binomial model for dose evaluation, and the entire design is built upon a set of fixed rules. 
		We show via simulation studies that the Ci3+3 design exhibits similar and comparable operating characteristics to more complex designs utilizing model-based inferences. 
		\yy Implementation of Ci3+3 for practical trials is simple for the first stage, where the up-and-down decisions may be carried out using a decision table based on the preselected escalation path and i3+3. The second stage is not simpler than model-based designs, however, since it also requires computation of posterior probabilities based on a Bayesian model. \jj
		We believe that the Ci3+3 design may provide an alternative choice to help simplify the design and conduct of combination dose-finding trials in practice.
	\end{abstract}
\end{titlepage}

\section{Introduction}
We consider combination dose-finding trials with two agents,    where    an ``agent'' refers to an investigational drug or therapy.  We assume that both agents have more than one candidate dose  under investigation.  For example, consider a trial with three dose levels of agent A and three levels of agent B, and suppose the first cohort receives the first levels of both agents A and B,    denoted by    dose combination $d_{11}$.    To be concise, we use ``DC'' to denote dose combination hereafter,    and    DC $d_{ij}$ refers to the dose combination in which agent A has dose level $i$ and agent B dose level $j$. 
Typically, phase I dose-finding trials aim to identify the maximum tolerated dose (MTD), which is defined as the highest dose with a toxicity probability close to or lower than a target rate $p_T$, say $p_T=0.3$ or $1/6$. With a single agent, a unique MTD is assumed to exist due to  the  assumption that toxicity increases  monotonically  with dose.  Therefore,   the order of toxicity probabilities across the ascending doses in a single-agent trial is completely known, referring to as the complete order   \citep{2014Phase}.  However, in dose combination trials with two agents, rather than a unique MTD, there  may  exist a set of MTD combinations, or MTDCs, as multiple DCs may satisfy the defition of MTD. For example,  both DCs $d_{23}$ and $d_{32}$ may be the highest DCs with a toxicity probability lower than $p_T$. Unlike the single-agent cases where the  order of all the doses is known, here we do not know the order of the    true    toxicity probabilities for $d_{23}$ and $d_{32}$. This is because $d_{23}$ has a lower dose for agent A than $d_{32}$ but a higher dose for agent B. In general, we do not know the  order of the DCs $d_{i_1, j_1}$ and $d_{i_2, j_2}$, if $(i_1 - i_2)(j_1 - j_2) < 0.$ This phenomenon is called  the partial ordering of doses  \citep{wages2011continual}. 

Due to  the issue of partial ordering,   dose escalation and de-escalation decisions for combination trials are more complex than those for single-agent trials. Specifically, escalation or de-escalation from a DC usually has two options, by increasing or decreasing the dose level of one of the two agents, respectively. Note that   simultaneous increment of  the dose levels  for   both agents is usually not allowed in a practical trial due to the ``no dose skipping'' rule    enforced in practice.        Therefore,    how to develop an    efficient dose-finding algorithm has been one of the focal questions   for   combination dose-finding trials.

A traditional approach to this problem is to    preselect    a subset of DCs with a    complete    order,   and   apply a single-agent design 
along the   selected DCs. For instance, in a  $three \times three$ dual-agent trial, a selected subset of DCs    with complete order may be    given by
$$d_{11} \rightarrow d_{12} \rightarrow d_{13} \rightarrow d_{23} \rightarrow d_{33}.$$
Here $a \rightarrow b$ means that DC $b$ is assumed to be at least as toxic as $a$.    This approach down-shifts the two-dimensional dose-finding space into a one-dimensional space, and   is  taken by much of the early work in dual-agent trials, such as \cite{korn1993using} and \cite{kramar1999continual}. The   apparent   disadvantage of this approach is that it limits the number of DCs that can be   tested   and it   will   miss   any   promising DCs located outside of the   selected set.   The waterfall design \citep{zhang2016practical} extends this approach and splits the entire DC space into multiple subsets, taking sub-trials within them. 
Another type of approaches \citep{conaway2004designs,wages2011continual} lays out multiple possible complete orders of the dose-toxicity relationship and decides which order is the most likely.    
\cite{mander2015product}   propose the product of independent beta probabilities dose escalation (PIPE)  design and  divide   the 
DC space into two parts by a boundary, avoiding directly modeling the DC ordering.  DCs below the boundary are assumed to have toxicity less than the targeted toxicity level (TTL), and    those above are assumed to be higher    than TTL. PIPE  proceeds by finding  
the most likely    boundary,   which can be used to estimate the MTDCs. Recently,  \cite{lin2017bayesian} and \cite{pan2017statistical} extend the rules of single-agent dose-finding designs, BOIN \citep{2015Bayesian} and Keyboard \citep{yan2017keyboard} for combination dose-finding trials to realize the exploration of the entire two-dimensional space. 
Lastly, fully model-based approaches on the basis of escalation with over-dose control (EWOC) have also been extensively researched    for combo dose finding.    For example,    \cite{Tighiouart2014DoseFW, Tighiouart2017ABA}   propose   a parametric model for the dose–toxicity relationship and   apply   the EWOC rules in the dose escalation/de-escalation process.    

We introduce a rule-based design, Ci3+3,  aiming to establish a simple and practical    solution    for dual-agent combination dose-finding trials with comparable performance to more sophisticated model-based designs.     Motivated by the i3+3 design    (Liu et al., 2020),   
Ci3+3 utilizies a set of up-and-down rules in a two-stage  design,  
exploring    the candidate DCs to learn their toxicity profiles and    identify a    MTDC. A set of deterministic rules will decide the dose assignment of patient cohorts throughout the trial,    which makes them relatively simple to apply and comprehend in practice.    
Despite    the    lack of model-based inference, we will show that the Ci3+3 design 
exhibits   comparable operating characteristics with more sophisticated designs in the reported simulation studies based on a variety of scenarios. One possible explanation is the use of advanced decision rules that take into account practical safety consideration and the small sample size    nature    of dose-finding trials, both of which reduce the performance of model-based inference. 
\yy Implementation of Ci3+3 for practical trials is simple for the first stage, where the up-and-down decisions may be carried out using a decision table based on the preselected escalation path and i3+3. The second stage is not simpler than model-based designs, however, since it also requires computation of posterior probabilities based on a Bayesian model. \jj

The remainder of the paper is organized as follows.    Section 2  provides  a brief review of the i3+3 design. In Section 3, we introduce the Ci3+3 design  and the corresponding dose-finding algorithm. In Section 4, we provide a hypothetical trial example. The operating characteristics of the Ci3+3 design are shown with two simulation studies in Section 5.  We end with a discussion and conclusion in Section 6.

\section{Review of i3+3 design}
We first review the i3+3   design and its   decision rules, upon which the Ci3+3 design is anchored. The i3+3 design assumes that patients are enrolled and assigned to doses in cohorts. The design requires input of the target toxicity rate $p_T$ and the equivalence interval EI $=[p_T-\epsilon_1, p_T-\epsilon_2]$, where $\epsilon_1$ and $\epsilon_2$ are two small positive fractions, such as 0.05. According to i3+3, any doses with toxicity probabilities within the EI are considered as the MTD. During the trial, suppose dose $d$ is currently used to treat patients. After the toxicity outcomes of the most recent cohort of patients at dose $d$ are observed, the i3+3 design identifies an appropriate dose for the next cohort of patients according to  a   simple algorithm    (Table \ref{tab:i3+3}   in Appendix A)    based on the observed data $(y_d, n_d)$ at the current dose $d$, where $n_d$ is the number of patients treated and $y_d$ the number of patients with dose limiting toxicity (DLTs). All the decisions    corresponding to possible $y_d$ and $n_d$ values can be pretabulated in advance, allowing investigators for examination before the trial starts. See Figure \ref{fig:i3+3-dtab} in Appendix A for an illustration.

\section{The Proposed Ci3+3 Design}
Similar to 
i3+3, the Ci3+3  design also  defines an equivalence interval (EI) $[p_T - \epsilon_1, p_T + \epsilon_2]$ using the target probability of toxicity $p_T$ and two small fractions $\{\epsilon_1, \epsilon_2\}$. Ci3+3 also applies 
up-and-down decision  rules     based on the relationship between  the observed toxicity rate and the EI. 

In general, for a dual-agent dose-finding trial, suppose $I$ dose 
levels of agent A and $J$    levels    of agent B are tested. 
   Let $d_{ij} = (d_i^A, d_j^B)$    be the actual  doses  of agents A and B for DC $(i,j)$, e.g. (0.3 mg/ml, 0.5 mg/ml). 
And denote $p_{ij}$ the probability of toxicity at $d_{ij}$.  
Assume toxicity is monotone    non-decreasing    with dose levels. Mathematically, this means that $p_{ij} \le p_{i,j+1}$ and $p_{ij} \le p_{i+1,j}$.

The proposed Ci3+3 design consists of two stages, with the first stage    aiming    for rapid escalation through a    ``escalation path'' (EP),    described in detail in $\S$\ref{sec:stage1}, and the second stage for    expansive exploration of    the DC space in $\S$\ref{sec:stage2}. The    Ci3+3    design assumes that patients are enrolled and assigned to DCs in cohorts,    i.e.,    the next cohort of patients may not be enrolled until toxicity outcomes from the previous cohort of patients have been fully observed.   The cohort size is set at three by default, but can be modified by investigators.  

\subsection{Stage I: Run-in Stage} \label{sec:stage1}
\begin{figure}[!htbp]
	\centering \includegraphics[width=0.4\textwidth]{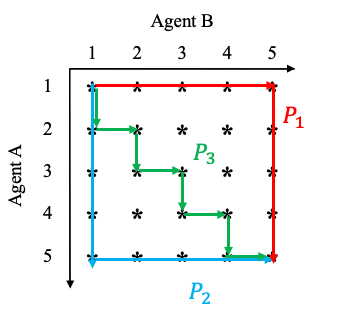}
	\caption{   Three examples of escalation paths (EPs) for Ci3+3.    Two    agents are to be tested, each with five dose levels.    EP $P_1$ represents a pathway in which the    DCs    firstly escalate    along    the dose level of agent B and then agent A.  EP $P_2$ represents the exact opposite of  $P_1$ , in which the    DCs    firstly escalate    along    the    dose    level of agent A and then agent B. EP $P_3$ alternates the escalation of    dose    level of agents A and B sequentially.    } \label{fig:Ci3+3-path}
\end{figure}

In Stage I, the Ci3+3 design follows the  i3+3    decision rules   and escalates along a prespecified EP to explore the DC space quickly. 
   An EP is   a subset of DCs with a complete order.  
Figure \ref{fig:Ci3+3-path} shows   three optional EPs,  denoted as $P_1$, $P_2$ and $P_3$, for a trial with five    dose levels of both agents A and B.    All three EPs contain DCs with  a  complete order,     given by   
$$P_1: \quad \left\lbrace d_{11} \rightarrow d_{12} \rightarrow d_{13} \rightarrow d_{14} \rightarrow d_{15} \rightarrow d_{25} \rightarrow d_{35} \rightarrow d_{45} \rightarrow d_{55} \right \rbrace,$$
$$P_2: \quad \left\lbrace d_{11} \rightarrow d_{21} \rightarrow d_{31} \rightarrow d_{41} \rightarrow d_{51} \rightarrow d_{52} \rightarrow d_{53} \rightarrow d_{54} \rightarrow d_{55} \right \rbrace,  \text{and}$$   
$$P_3: \quad \left\lbrace d_{11} \rightarrow d_{21} \rightarrow d_{22} \rightarrow d_{32} \rightarrow d_{33} \rightarrow d_{43} \rightarrow d_{44} \rightarrow d_{54} \rightarrow d_{55} \right \rbrace.$$

\noindent 
The i3+3 design (Liu et al. 2020)  is used to  guide the first-stage dose escalation due to its simplicity and consistency with the upcoming Stage II decisions. However, it needs not to be i3+3 as many other single-agent designs are available.    The determination of EP is discussed later in \S 3.3. 

   Stage   I   starts by treating the first cohort of patients at DC $d_{11}$. When the toxicity outcomes of the previous cohort of patients are observed, the i3+3 design as described in Table \ref{tab:i3+3} is applied to guide the dose selection for the next cohort of patients. In particular, Stage I continues as long as the decision is $E$, escalate, in which case the next cohort of patients is treated at the higher DC along the EP. When the i3+3 decision is not $E$, but $S$ stay or $D$ de-escalate, Stage I ends and Stage II starts. Therefore, Stage I allows quick escalation along a selected EP until escalation is no longer possible. The last DC treated in Stage I is the starting DC of Stage II. If the escalation in Stage I reaches the highest DC, Stage I ends and Stage II starts at the highest DC.   



\subsection{Stage II: Adaptive Stage} \label{sec:stage2}
   In Stage II, the full space of DCs is explored. Stage II starts at the last DC of Stage I and continues to assign  the next cohort of patients using an algorithm extending the rules in the i3+3 design.    

Suppose DC    $d_{ij}$    is currently used in the trial to treat patients,    at which    $y_{ij}$ patients have experienced DLT out of $n_{ij}$ enrolled patients. 
   Stage II    applies    the same    up-and-down decisions   $E$, $S$ or $D$   to decide the DC for the next cohort of patients.   However,   in  the two dimensional DC space,   since   only   partial order of DCs is known, each one of the three up-and-down decisions may lead to multiple candidate DCs. For example, a decision $E$ from DC $d_{ij}$ may lead to $d_{i+1, j}$ or $d_{i, j+1}$ as the next DC in the trial. Note that, DC $d_{i+1, j+1}$ is not considered in $E$ since in practice increasing dose levels in both agents is typically not allowed due to safety concerns. For the same reason, decisions $S$ and $D$ may imply more than one DC for the next cohort. In other words, the up-and-down decisions in Stage II  point to a set of candidate DCs, rather than a single one. We first list the set of candidate doses corresponding to each of the three decisions next, assuming $d_{ij}$ is the current DC used in the trial.       

\begin{figure}[!htbp]
	\centering \includegraphics[width=0.4\textwidth]{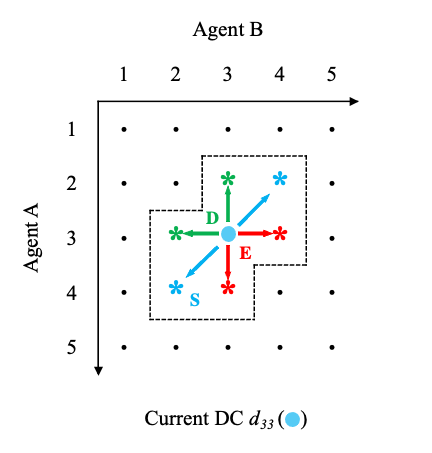} 
	\caption{   An example of the adjacent   candidate DCs. The dashed box contains the candidate DCs, 
      which      correspond to three candidate sets of DCs for Stage II of Ci3+3:    red stands for the candidate DCs   for 
	  $E$, blue for $S$, and green for $D$.} \label{fig:Ci3+3-admis}
\end{figure}

We start by defining a distance of two DCs. For a DC $d_{ij}$, we call a DC $d_{kl}$ a “$M\circ$DC"   if the maximum value of differences between $i$ and $k$, and between $j$ and $l$, is equal to $M$, $M=1, 2, ...$.   Mathematically, this means that   $M = \max (|i-k|,|j-l|)$.   Let $\Omega_{ij}^{(E)},\Omega_{ij}^{(S)}$,    and $\Omega_{ij}^{(D)}$ denote   the adjacent  candidate  sets of DCs    for the current DC $d_{i,j}$ for decision escalation ($E$), stay ($S$) and de-escalation ($D$), respectively. They are defined to be 
\begin{align*}
\Omega_{ij}^{(E)} &=\left\{d_{i'j'} \mid 1 \le i' \le I, 1 \le j' \le J, |i'-i| \le 1, |j'-j| \le 1, (i'-i)+(j'-j)=1\right\}, \\
\Omega_{ij}^{(S)} &=\left\{d_{i'j'} \mid 1 \le i' \le I, 1 \le j' \le J, |i'-i| \le 1, |j'-j| \le 1, (i'-i)+(j'-j)=0\right\}, \\
\Omega_{ij}^{(D)} &=\left\{d_{i'j'} \mid 1 \le i' \le I, 1 \le j' \le J, |i'-i| \le 1, |j'-j| \le 1, (i'-i)+(j'-j)=-1\right\}.
\end{align*}
 In words, the three adjacent candidate sets are the subsets of  $1\circ$DCs to $d_{ij}.$  Figure \ref{fig:Ci3+3-admis} gives an example, where the current DC is $d_{33}$,  $\Omega_{33}^{(E)}=\left\{d_{34}, d_{43} \right\}$, $\Omega_{33}^{(S)}=\left\{d_{24}, d_{33}, d_{42} \right\}$,  and $\Omega_{33}^{(D)}=\left\{d_{23}, d_{32} \right\}$. 

%

Once the adjacent candidate sets are determined,  Stage II of Ci3+3   uses a dose-finding algorithm to determine an appropriate  DC for the next cohort of patients continuously. The main idea is to apply the up-and-down decisions and select a candidate DC based on a simple working Bayesian model that produces posterior probabilities for a DC to be the MTDC as a measure for the DC's desirability. The algorithm is provided below.

\bigskip

\paragraph*{Stage II algorithm:  }
Suppose DC $d_{ij}$ is being used to treat patients in the trial. Based on the observed data $(y_{ij}, n_{ij})$ we want to identify an appropriate DC to treat the next cohort of patients. Let us call this DC ``the next DC''. We first summarize the main ideas of the algorithm to facilitate the upcoming discussion. Also, we call a DC {\it orderless} to the adjacent candidate set $\om{X}$ if the order of the toxicity probability between the DC and any DC in $\om{X}$ is unknown, $X \in \{E, S, D\}$. Note that all DCs within the set $\om{X}$ are orderless. In contrast, we call a DC {\it higher} or {\it lower} than another DC or a set of DCs, if its toxicity probability is higher or lower than the DC or all the DCs in set, respectively. Below we make two remarks before introducing the dose-finding algorithms.

\begin{description}
  \item[] First, we determine the up-and-down decisions $\cA_{ij} \in \{E, S, D\}$ from the i3+3 design based on the observed data $(y_{ij}, n_{ij})$ at the current DC $d_{ij}$. The decision $\cA_{ij}$ indicates that the next DC may be from the adjacent candidate set $\om{\cA_{ij}}$. That is, if $\cA_{ij}$ equals $E$, $S$, or $D$, the next DC will be selected from the adjacent candidate set $\om{E}$, $\om{S}$, or $\om{D}$, respectively. 
    For example, if $\cA_{ij}$ equals $E$, then we consider DCs in $\om{E}$, which are the $1\circ$DCs higher than $d_{ij}$. 
  
  \item[] Second, we consider  two conditions that would  encourage exploration of the DC space. Let $d_{kl} $ denote a $1\circ$DC in the adjacent candidate set $\om{\cA_{ij}}$ for the current DC $d_{ij}$.  Define two conditions as follows.  
		\begin{itemize}
		\item[] {\bf Condition 1} all the DCs $d_{kl}$'s in the adjacent candidate set $\om{\cA_{ij}}$ have already been tested, and
		\item[] {\bf Condition 2} the corresponding decision is  $\cA_{kl} = S$ for all $d_{kl} \in \om{\cA_{ij}}$.
		\end{itemize}
	     When conditions 1 \& 2 are satisfied, instead of selecting a DC from $\om{\cA_{ij}}$, we will consider the orderless and untested $1\circ$DCs to $\om{\cA_{ij}}$ (i.e., $1\circ$DCs to each DC in the adjacent candidate set) for future patients. This means assigning patients to potential $2\circ$DCs.    	

 \end{description}

 While these rules may sound complicated, they are needed to increase the efficiency of dose finding. Finally, we will need  the posterior probability of belonging to EI, defined as $\xi_{ij} = Pr\{p_{ij} \in EI \mid y_{ij},n_{ij} \}. $

Here, the posterior probability is calculated using a working Bayesian model that assumes the binomial likelihood for the observed data $\{y_{ij}, n_{ij}\}$ on DC $d_{ij}$ and a beta prior distribution, $Beta(1,1),$ for $p_{ij}$. That is, the posterior distribution of $p_{ij}$ is $Beta(1+y_{ij},1+n_{ij}-y_{ij})$ given $y_{ij}$ DLTs out of $n_{ij}$ patients at DC $d_{ij}$. A higher value of $\xi_{ij}$ means the DC $d_{ij}$ is more likely to be the MTDC by its definition. Below we provide the step-by-step dose-finding algorithm for Stage II.

In words, the algorithm in general prefers to assign future patients to the DC in the candidate set with the largest posterior probabilities being the MTDC, except for a few special cases in the remarks above. In those cases, untested DCs are preferred to encourage exploration of new DCs.

\clearpage

\begin{algorithm}[htbp]
  \caption{Stage II algorithm of Ci3+3	  }
  \label{stageIIrules}
  \begin{algorithmic}

  \STATE Suppose DC $d_{ij}$ is currently administered to patients enrolled in the trial and the decision for the next cohort of patients is $\cA_{ij} \in \{E,S,D\}$.
 
	\IF {$n_{kl} > 0$ and $\cA_{kl}=S$, $\forall d_{kl} \in \om{X},$ i.e., each DC in $\om{X}$ has been tested and if i3+3 were applied to make a decision based on the observed data at the DC, the decision would be $S$,}
		\STATE denote $\om{X2}$ the set of all the orderless $1\circ$DCs to $\om{X}$ that are untested; mathematically, \\$\om{X2} = \{d_{pq} \mid \forall d_{kl} \in \om{X}, 1 \le p \le I, 1 \le q \le J,  |p-k| \le 1, |q-l| \le 1, (p-k)+(q-l)=0, n_{pq} =0 \},$

		\IF {$\om{X2} \ne \emptyset$, i.e., there exists untested $1\circ$DCs to $\om{X}$,}
			\STATE {\bf the next DC is randomly selected among those DCs $d_{pg} \in \om{X2}$}
		\ELSE
			\STATE {\bf the next DC is $\arg\max_{d_{kl}\in \om{X}} \xi_{kl}$;}
			\STATE In English, the next DC is $d_{kl}$ that has the largest posterior probability of belonging to EI in the adjacent candidate set $\om{X}$. 
		\ENDIF
	\ELSE
		\STATE {\bf the next DC is $\arg\max_{d_{kl}\in \om{X}} \xi_{kl}$;}
	\ENDIF
 
\end{algorithmic}
\end{algorithm}

    In Appendix B we introduce an optional rule in addition to Algorithm \ref{stageIIrules}. The optional rule encourages exploration of new DCs but may result in slightly reduced safety in simulation. As a tradeoff, the optional rule allows more DCs to be explored. See Appendix B for details. \jj 

\subsection{ Practical  Rules and MTDC selection}
We consider a rule for safety, which has been widely used in dose-finding literature.

\paragraph{DC Exclusion and Early Stopping Rule} \label{sec:dc_exclude} 
If DC $d_{ij}$ is considered with excessive toxicity,    the DC    and all higher DCs with known    order $\left\{d_{i'j'}\mid i \le i' \le I, j \le j' \le J\right\}$ are excluded from the trial and never used again in the remainder of the trial. 
We   deem   DC $d_{ij}$  overly toxic  
if   
$$Pr\left\{ p_{ij}>p_T \mid y_{ij},n_{ij} \right\}>\xi,$$
where  $n_{ij} \ge 3$   and the threshold $\xi$ is close to 1, say 0.95. And $Pr\left\{ p_{ij}>p_T \mid y_{ij},n_{ij} \right\}$ is   calculated under the beta distribution,    $Beta(\alpha_0+y_{ij},\beta_0+n_{ij}-y_{ij})$,   with $\alpha_0=\beta_0=1$.
If $d_{11}$ is deemed  overly toxic,  
the trial is terminated.  

\vskip 0.2in 

Next, we propose the statistical inference for MTDC selection after all patients in the trial have been enrolled and their data observed.

\paragraph{MTDC Selection}  \label{sec:terminate-and-MTDC-sel}

The trial stops either if $d_{11}$ is overly toxic or when the prespecified maximum sample size $N$ is reached. If $d_{11}$ is overly toxic, no MTDC is selected. Otherwise, we select  a  MTDC based on the following procedure.
First of all, we assume that the prior for each $p_{ij}$ follows an indepedent $beta(0.005,\ 0.005)$, 
and the posterior distribution for each $p_{ij}$ is given by $beta(0.005 + y_{ij},\ 0.005 + n_{ij} - y_{ij})$.
We then estimate $p_{ij}$ by calculating the posterior mean of each DC, 
which is given by ${(y_{ij} + 0.005)} / {(n_{ij} + 0.01)}$,
and perform a bivariate isotonic regression   \citep{biviso}   on the posterior means to meet the monotonic dose-toxicity assumption.
Denote the isotonic-transformed posterior means $\hat p_{ij}$ for all the DCs.
Next, we eliminate DCs at which the number of enrolled patients is less than or equal to 3, (i.e, $n_{ij} \le 3$)
and  DCs that are excessively toxic (i.e, $Pr\{p_{ij} > p_T | y_{ij}, n_{ij}\} > \xi$ or $\hat p_{ij} > p_T + \epsilon_2$). These elimination improve the operating characteristics of the designs by weeding out DCs with little information or with potential excessive toxicity.   
Finally, we select the DC   for which the $\hat p_{ij}$ is the closest to the target rate $p_T$  as the MTDC .
When there are ties for $\hat p_{ij}$'s with the same index $i$ or $j$, we select the highest DC (largest $i$ or $j$)  among the tied DCs if   $\hat p_{ij} < p_T$,
or the lowest DC (smallest $i$ or $j$)  if   $\hat p_{ij} > p_T$, as the MTDC. If the   tied $\hat p_{ij}$'s  have different $i$ and  $j$, we randomly pick one as the MTDC.

The procedure described above selects one DC as the MTDC. If desired, the selection procedure may be modified to report multiple MTDCs. In the literature, some designs (such as \citealp{mander2015product} and \citealp{zhang2016practical}) allow multiple MTDCs to be selected, while the other designs (such as \citealp{lin2017bayesian} and \citealp{pan2017statistical}) limit the MTDC selection to one DC.

\vskip 0.2in
Lastly, we provide a discussion on the EP selection.

\paragraph{EP Selection} 

 Taking into account that investigators have different prior preclinical and clinical information on the toxicity of the two-agent combinations,    we recommend two   methods   for EP selection in Stage I.    
\begin{itemize}
	\item[--]    {\bf Method 1:}   If one agent is less well understood (e.g., a new therapeutics) than the other agent (e.g., a well-established treatment) in its dose response, choose an EP that encourages escalation along the dose levels of the agent less well understood. For example, in Figure \ref{fig:Ci3+3-path}  EP 2  may be selected if agent A is less well understood.

	
	
	\item[--]    {\bf Method 2:} If little prior information is known on the toxicity of both   agents,   EP $P_3$ in Figure \ref{fig:Ci3+3-path} is recommended as it has    shown    desirable    operating characteristics    based on our simulation. This EP alternates the dose level of either agent when escalating.    

\end{itemize}

\section{A Trial Example}
 
\begin{figure}[H]
  \centering \includegraphics[width=\textwidth]{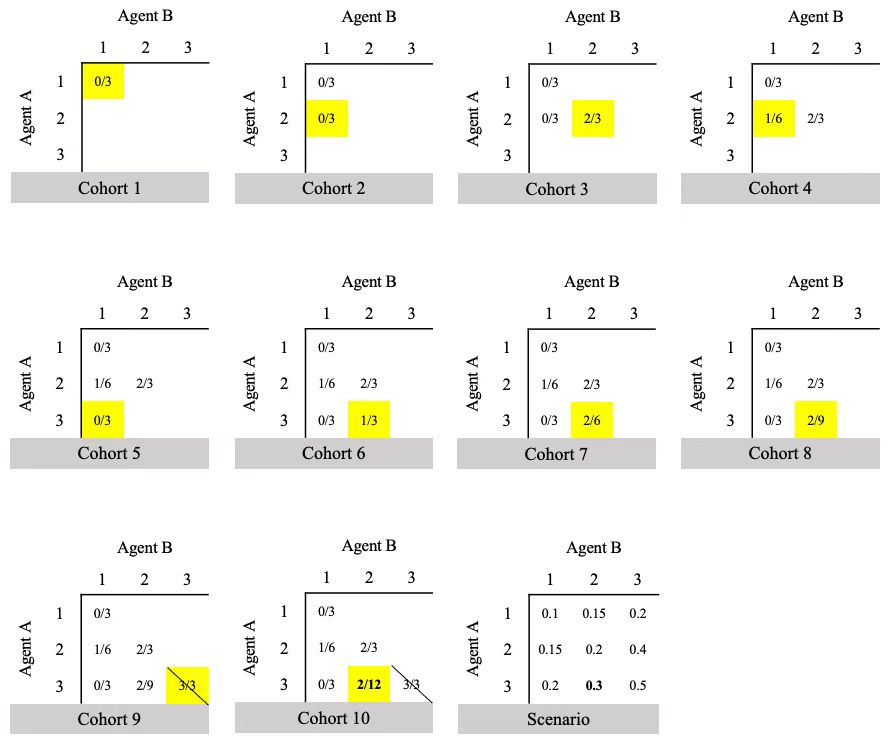}
  \caption{   A trial example of the Ci3+3 design. There are 10 cohorts in this example. The DCs used to treat each cohort are highlighted in yellow. The values at each DC are the number of patients with DLTs and the number of patients treated (i.e., $y_{ij}$/$n_{ij}$), respectively. The last panel shows the scenario truth of this example, and the values at each DC indicate the true toxicity probabilities. In Cohort 10, DC $d_{32}$ is selected as the MTDC, and its data 2/12 is in bold font. }
  \label{fig:ci3-example}
\end{figure}

   To illustrate the algorithm of the Ci3+3 design, here we give a hypothetical trial example (Figure \ref{fig:ci3-example}). In this example, a total of 30 patients is enrolled in cohorts, each cohort with 3 patients. 

In stage I, we use $P_3$ as the EP for dose escalation and use i3+3 decisions to guide the dose selection process. The first cohort of patients are enrolled at DC $d_{11}$. There are 0 DLTs in this cohort, and so the decision is E. Cohort 2 is assigned to $d_{21}$, and the decision is again E; as a result, Ci3+3 assigns Cohort 3 to $d_{22}$. At $d_{22}$, 2 of 3 patients experience DLT, and the Ci3+3 decision is D. Since the decision is not E, stage I ends and stage II starts with $d_{22}$ as the starting DC. 

When stage II starts at DC $d_{22}$, the Ci3+3 decision is D. There are two DCs in the candidate set, that is $\Omega_{22}^{(D)}$ =  $\{ d_{21}, d_{12} \}$. Therefore, Ci3+3 assigns Cohort 4 to $d_{21}$ because $d_{21}$ has a higher posterior probability of belonging to the EI than $d_{12}$. After Cohort 4's outcomes are observed, which have 1 DLT, there is a total of 1 DLT out of the 6 patients treated at $d_{21}$ (3 patients from previous Cohort 2). Therefore, the decision is E, and Ci3+3 assigns Cohort 5 to $d_{31}$, which has a higher probability  of belonging to the EI than the other candidate DC $d_{22}$. At $d_{31}$, there are 0 out of 3 patients experiencing DLTs, and so the decision is E. Since there is only one candidate DC $d_{32}$ in $\Omega_{31}^{(E)}$, Ci3+3 assigns Cohort 6 to $d_{32}$.

From Cohort 6 to Cohort 8, patients continue to be enrolled at $d_{32}$. Now there are 2 DLTs out of 9 patients at the current DC $d_{32}$, so the Ci3+3 decision is E. Since there is only one candidate DC in $\Omega_{32}^{(E)}$, Ci3+3 assigns Cohort 9 to $d_{33}$.  There are 3 DLTs out of 3 patients in Cohort 9, and so the decision is D and $d_{33}$ is excluded from the trial by the DC Exclusion Rule. Cohort 10 is assigned to $d_{32}$, which has a higher probability of belonging to the EI than $d_{23}$.  At the end of the trial, Ci3+3 selects $d_{32}$ as the MTDC.

\section{Simulations}
\subsection{Simulation    Study    1}
\paragraph{Simulation Setup}
We   assess   the operating characteristics of the Ci3+3 design with comparison   to the PIPE \citep{mander2015product}, waterfall \citep{zhang2016practical}, BOIN combo \citep{lin2017bayesian}    and Keyboard combo \citep{pan2017statistical}   designs using   seven scenarios in Table \ref{tab:sim1} from \cite{braun2013generalized}.  We also add one scenario (Scenario 8) with toxicity that plateaus at the end.  For each scenario, $1,000$ trials are simulated using the true  probabilities  of toxicity in the scenario and applying each of  the candidate designs. Each trial targets a toxicity probability $p_T=30\%$, with a maximum sample size of 96. Each trial uses   three   patients per cohort and starts at DC $d_{11}$. 
For the PIPE design, we take the closest DC to define the candidate dose set,  choose the candidate dose with the smallest current sample size, and use a weak prior, $a_{ij}=b_{ij}=0.5$,   for all DC $d_{ij}$.  
For the waterfall design, we use  14  cohorts for the first subtrial and  6   cohorts for each of the next three subtrials.
%
%
  
We  apply  the default settings of the “get.oc.comb.kb" function in the R package “Keyboard" for the Keyboard combo design, and the default settings of the “get.oc.combo" function in the R package ``BOIN" for the waterfall and BOIN combo designs, except that we do not allow waterfall to stop a subtrial early.   This improves the performance of waterfall although it increases the average sample size across simulations.  
  
For the Ci3+3 design, we  set   $\epsilon_1=\epsilon_2=0.05$, that is EI=[0.25,0.35], and a weak prior, $a_{ij}=b_{ij}=0.05$, for the MTDC selection. For the PIPE deisn, we terminate a trial if $d_{11}$ is excessively toxic according to the Early Stopping rule in  \S \ref{sec:dc_exclude}   to make comparison more fair since all other designs use it.  We choose EP $P_3$ by default for the Stage I dose finding.  

\paragraph{Simulation Results}
We characterize the true MTDCs as any DC with a toxicity probability inside the EI. If no    DCs    satisfy the criterion, the true MTDC is the highest DC among those with true toxicity probabilities $p_{ij}<p_T$. If no MTDC could still be identified    based    on the above two    criteria,    the paticular scenario does not have a true MTDC, which means the   correct  decision is to select no DC as the MTDC.  

To assess the performance of a dual-agent dose-escalation design, we    report    the following  operating characteristics  about the MTDC selection,
\begin{enumerate}
	
	\item    The    percentage of correct selection (PCS), defined as the percentage of the trials where at least one    true    MTDC    is selected.    If there are no true MTDCs   in the scenario, the correct decision is no selection, and PCS is equal to the percentage of the trials in which no DCs are selected.   
	  
	\item The percentage of  overdose selection (POS), defined as the percentage of the trials where at least one DC above the    true MTDC(s) is selected.    
	
	\item The percentage of underdose selection (PUS), defined as the percentage of the trials where at least one DC below the    true    MTDC(s)    is selected.   
	
	\item Average number of DCs   selected across all the trials (AvgNsel). This is needed since some designs like waterfall may select $> 1$ MTDCs for one trial.
\end{enumerate}
We also report    the following statistics about the patients allocation,
\begin{enumerate}
	\item    Correct allocation (CA):    Average number of patients allocated at the true MTDC(s).
	\item    Over allocation (OA):    Average number of patients allocated at DCs above the true MTDC(s).
	\item    Under allocation (UA):    Average number of patients allocated at DCs below the true MTDC(s).
	\item    Total:    Average total number of enrolled patients in the trials.
\end{enumerate}
 Apart from the aforementioned operating characteristics, in Appendix D, we also report the Accuracy Index \citep{cheung2011dose} based on the selected dose level: 
$$
\text{Accuracy Index} = 1 - n_{D1} \times n_{D2} \times \frac {
	\sum_{i=1}^{n_{D1}}\sum_{j=1}^{n_{D2}}\rho_{ij} \times \text{Percentage of selecting dose level} \ d_{ij}
	} {\sum_{i=1}^{n_{D1}}\sum_{j=1}^{n_{D2}}\rho_{ij}}
$$
and the assignment index based on the percentage of subjects assigned:
$$
\text{Assignment Index} = 1 - n_{D1} \times n_{D2} \times \frac {
	\sum_{i=1}^{n_{D1}}\sum_{j=1}^{n_{D2}}\rho_{ij} \times \text{Percentage of subjects assigned to dose level} \ d_{ij}
	} {\sum_{i=1}^{n_{D1}}\sum_{j=1}^{n_{D2}}\rho_{ij}},
$$
where $\rho_{ij} = \left|  p_{ij}-p_{T}  \right|$ and  $p_{ij}$ is the true toxicity probability at dose level $d_{ij}$, $p_{T}$ is the target DLT rate, and $n_{D1}$ and $n_{D2}$ are the number of dose levels for agent A and agent B, respectively.

Figures \ref{fig:sim1_sele} and \ref{fig:sim1_allo} summarize the statistics for all the designs  and  
scenarios.   In general, the Ci3+3 design is comparable to other designs in PCS and CA.   
The waterfall design allows more than one DC to be selected as the MTDC, and so has a larger PCS. The Ci3+3 design has slightly lower POS, which means that Ci3+3 is more conservative and  safer  when selecting the MTDC.  However, the Ci3+3 design has slightly higher OA values.  The reason is that the Ci3+3 design encourages patients allocation to less explored DCs    during the dose-finding process.    And such an encouragement may result in assigning slightly more patients to DCs above the true MTDC.


\begin{table}[!htbp]
	\centering
	\caption{   True probabilities of toxicity (in percentage)    for the seven scenarios in \cite{braun2013generalized} with the true MTDCs shown in bold.}
	\begin{tabular}{|ccrrrrrrcrrrr|}
		\cline{1-13}
		&       & \multicolumn{4}{c}{Agent B}    &       &       &       & \multicolumn{4}{c|}{Agent B} \\
		\multirow{5}[0]{*}{Scenario 1} & Agent A & 1     & 2     & 3     & 4     &       & \multicolumn{1}{c}{\multirow{5}[0]{*}{Scenario 5}} & Agent A & 1     & 2     & 3     & 4 \\
		& 1     & 4     & 8     & 12    & 16    &       &       & 1     & 8     & 18    & \textbf{28} & \textbf{29} \\
		& 2     & 10    & 14    & 18    & 22    &       &       & 2     & 9     & 19    & \textbf{29} & \textbf{30} \\
		& 3     & 16    & 20    & 24    & \textbf{28} &       &       & 3     & 10    & 20    & \textbf{30} & \textbf{31} \\
		& 4     & 22    & \textbf{26} & \textbf{30} & \textbf{34} &       &       & 4     & 11    & 21    & \textbf{31} & 41 \\
		&       &       &       &       &       &       &       &       &       &       &       &  \\
		&       & \multicolumn{4}{c}{Agent B}    &       &       &       & \multicolumn{4}{c|}{Agent B} \\
		\multirow{5}[0]{*}{Scenario 2} & Agent A & 1     & 2     & 3     & 4     &       & \multicolumn{1}{c}{\multirow{5}[0]{*}{Scenario 6}} & Agent A & 1     & 2     & 3     & 4 \\
		& 1     & 2     & 4     & 6     & 8     &       &       & 1     & 12    & 13    & 14    & 15 \\
		& 2     & 5     & 7     & 9     & 11    &       &       & 2     & 16    & 18    & 20    & \textbf{22} \\
		& 3     & 8     & 10    & 12    & 14    &       &       & 3     & 44    & 45    & 46    & 47 \\
		& 4     & 11    & 13    & 15    & \textbf{17} &       &       & 4     & 50    & 52    & 54    & 55 \\
		&       &       &       &       &       &       &       &       &       &       &       &  \\
		&       & \multicolumn{4}{c}{Agent B}    &       &       &       & \multicolumn{4}{c|}{Agent B} \\
		\multirow{5}[0]{*}{Scenario 3} & Agent A & 1     & 2     & 3     & 4     &       & \multicolumn{1}{c}{\multirow{5}[0]{*}{Scenario 7}} & Agent A & 1     & 2     & 3     & 4 \\
		& 1     & 10    & 20    & \textbf{30} & 40    &       &       & 1     & 1     & 2     & 3     & 4 \\
		& 2     & \textbf{25} & \textbf{35} & 45    & 55    &       &       & 2     & 4     & 10    & 15    & 20 \\
		& 3     & 40    & 50    & 60    & 70    &       &       & 3     & 6     & 15    & \textbf{30} & 45 \\
		& 4     & 55    & 65    & 75    & 85    &       &       & 4     & 10    & \textbf{30} & 50    & 80 \\
		&       &       &       &       &       &       &       &       &       &       &       &  \\
		&       & \multicolumn{4}{c}{Agent B}    &       &       &       &   \multicolumn{4}{c|}{Agent B} \\
		\multirow{5}[1]{*}{Scenario 4} & Agent A & 1     & 2     & 3     & 4     &       &   \multicolumn{1}{c}{\multirow{5}[0]{*}{Scenario 8}} & Agent A & 1     & 2     & 3     & 4 \\
		& 1     & 44    & 48    & 52    & 56    &       &       &   1    &   1    &    2    &    3   &  4 \\
		& 2     & 50    & 54    & 58    & 62    &       &       &   2    &   4    &    10   &   15   &  20 \\
		& 3     & 56    & 60    & 64    & 68    &       &       &   3    &   6    &    15   &   \textbf{30}   &  36 \\
		& 4     & 62    & 66    & 70    & 74    &       &       &   4    &   10   &    \textbf{30}   &   38   &  40 \\

	    \cline{1-13}
	\end{tabular}
	\label{tab:sim1}
\end{table}
\newpage



\begin{figure}[!htbp]
	\begin{subfigure}{\textwidth}
		\includegraphics[width=\textwidth]{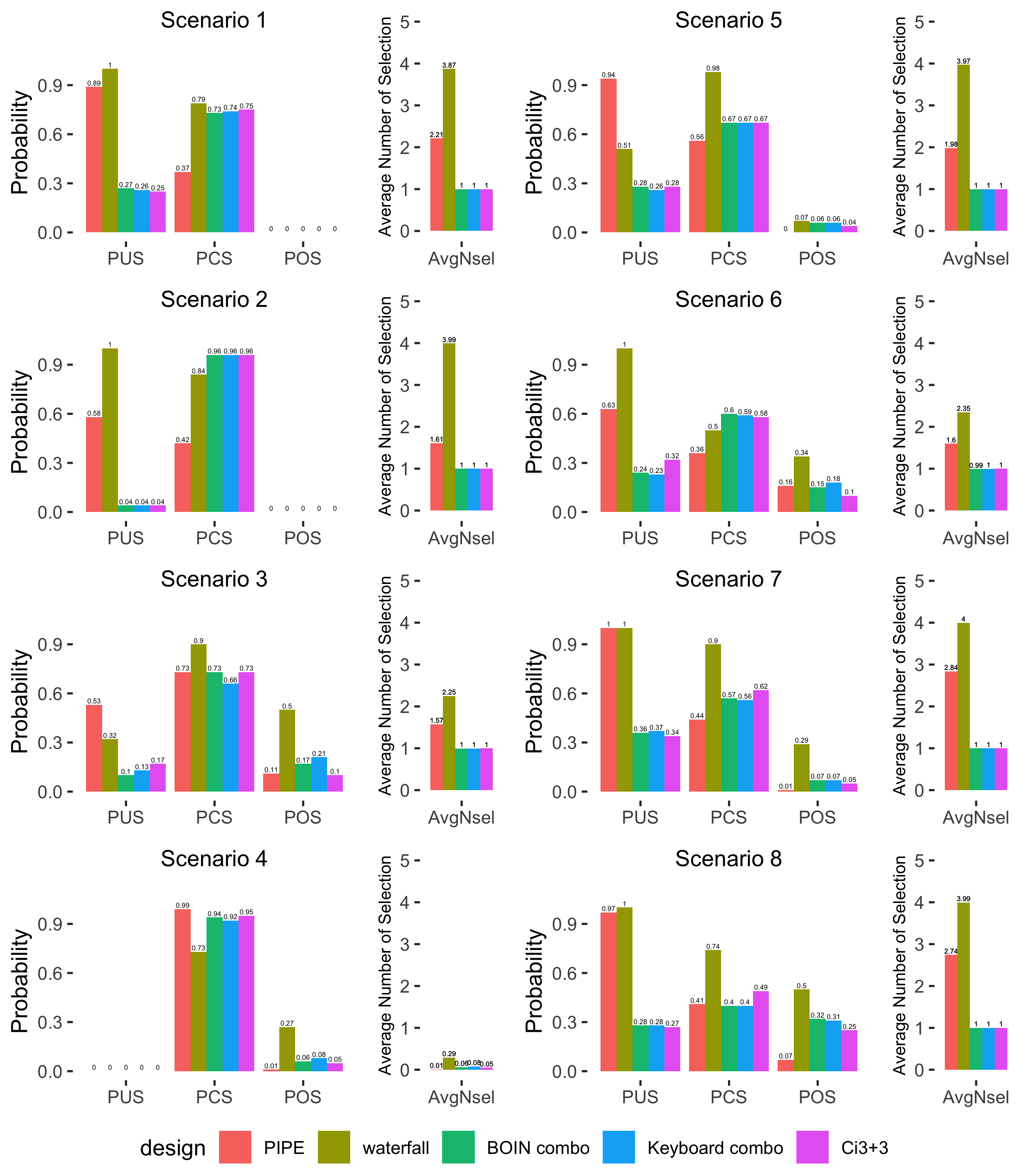}
		\subcaption{ DC selection (in probability) in  Simulation Study 1. The waterfall design allows more than one DC to be selected as the MTDC, and therefore the sum of PUS, PCS and POS may be greater than 1. }
		\label{fig:sim1_sele}
	\end{subfigure}
    \caption{Results of Simulation Study 1.}
\end{figure}%

\begin{figure}[!htbp]\ContinuedFloat
	\begin{subfigure}{\textwidth}
		\includegraphics[width=\textwidth]{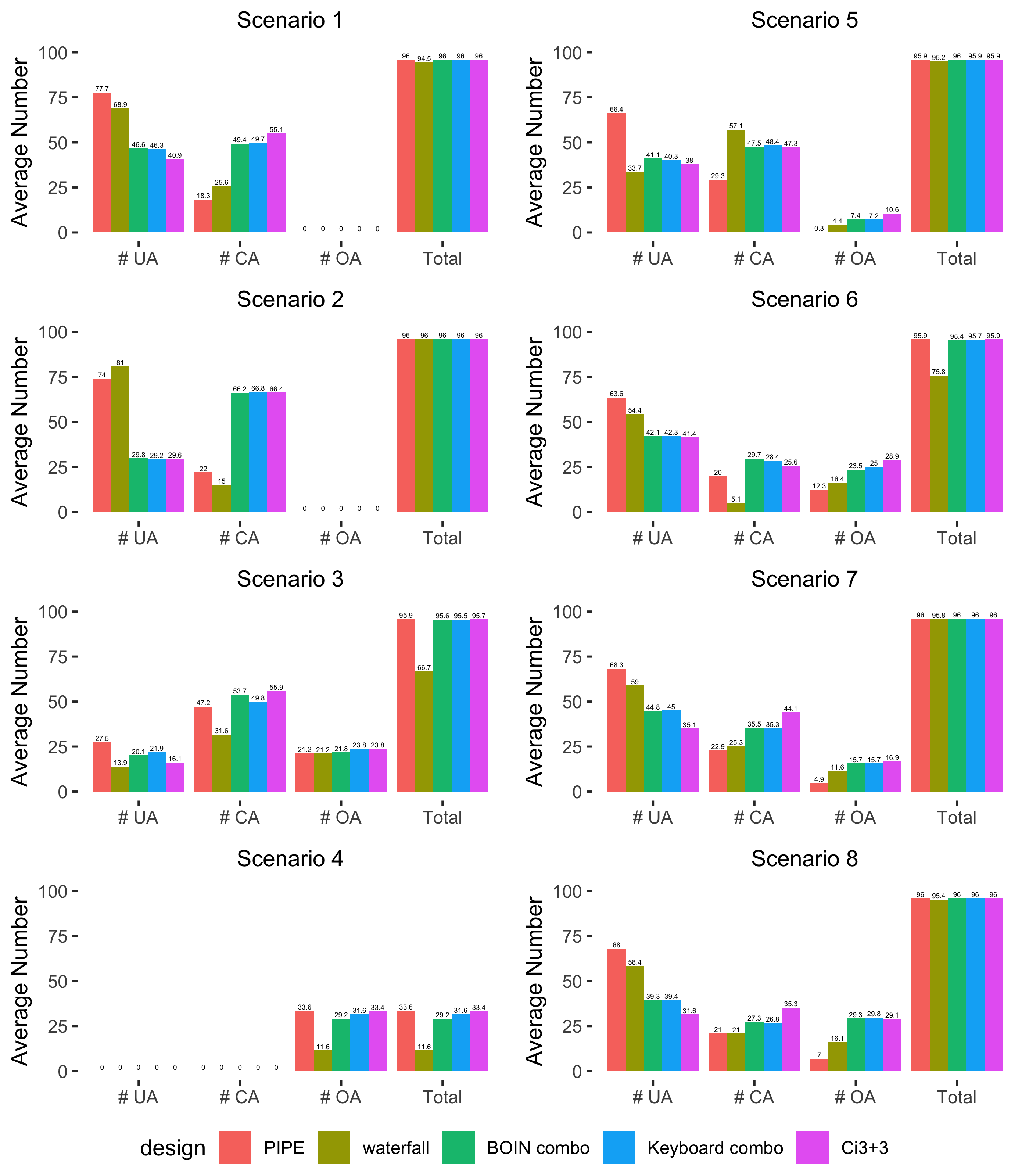}
		\subcaption{Patients allocation  for Simulation Study 1.   Reported are the average number of patients on each bar.    }
		\label{fig:sim1_allo}
	\end{subfigure}
	
	\caption{Results of Simulation Study 1.}
    \label{fig:sim1}
\end{figure}

\subsection{Simulation    Study    2}
\paragraph{Simulation Setup}
  We  conduct a larger     simulation study and    investigate the operating characteristics of the PIPE, waterfall, BOIN combo,    Keyboard combo    and Ci3+3 designs in 100 scenarios generated from the model introduced   in    \cite{neuenschwander2015bayesian}. According to this   dual-combination   model, the toxicity probabilities of the two-agent combinations can be calculated by the marginal toxicity probabilities and  an  interaction coefficient. 

Let $p_{A,i}$ and $p_{B,j}$ be toxicity probabilities of two single agents ascribed to $i$-th level of agent A and $j$-th level of agent B, respectively, for $i=1,2,\ldots,I$ and $j=1,2,\ldots,J$. In the special case of no interaction (independence), the single-agent toxicities fully determine the toxicities of combinations. For dose combination  $d_{ij}$,   the probability of no DLT is $(1-p_{A,i})(1-p_{B,j})$. Under independence, let $p_{ij}^0$ be the toxicity probability of the combination  $d_{ij}$   when the two agents are independent, it is true that
\begin{equation*}
p_{ij}^0 = 1- (1-p_{A,i})(1-p_{B,j}) = p_{A,i} + p_{B,j} - p_{A,i}p_{B,j}. 
\end{equation*}
On the odds scale this is equivalent to
\begin{equation*}
odds_{ij}^0 = odds_{A,i} + odds_{B,j} + odds_{A,i} \times odds_{B,j}, 
\end{equation*}
 where $odds_{ij}^0=p_{ij}^0/(1-p_{ij}^0)$, etc. To allow interaction, one assumes  
\begin{equation*}
odds_{ij} = odds_{ij}^0 \times g(\eta,i,j). 
\end{equation*}
   The same interaction  $g(\cdot)$ is used    for all dose combinations, i.e., $g(\eta,i,j)=\exp(\eta)$. Different values of $\eta$ represent different relationship between the two agents, namely, interaction coefficient. Lastly, we have toxicity probabilities for   DC $d_{ij}$    through $p_{ij}=\frac{odds_{ij}}{1+odds_{ij}}$.

Assuming that the marginal toxicity probabilities of two single agents    are given by     one of five different scenarios and there are four possible values of $\eta$. We then arrive at $\binom{5}{2} \times 4 = 100$ unique scenarios of toxicity probabilities for two-agent combinations. See Appendix C for details.   

Other settings of the    five    designs are the same as Simulation Study 1. For the waterfall design, there are 14 cohorts pre-allocated to the first subtrial and 6 cohorts to each of the next three subtrials.

\paragraph{Simulation Results}
We report the   MTDC selection and patient allocation statistics in Table \ref{tab:sim2_sel_and_alloc}, averaged over 100 scenarios.   
Ci3+3 performs comparable to other designs, with comparable PCS and CA, slightly smaller POS but higher OA.

  

\begin{table}[!htbp]
	\centering
	\caption{Simulation results for Simulation Study 2.}
	Average MTDC selection in 100 scenarios.
	\begin{tabular}{c|ccc|c}
		\hline
		\textbf{Design} & \textbf{PUS(s.d.)} & \textbf{PCS(s.d.)} & \textbf{POS(s.d.)} & \multicolumn{1}{c}{\textbf{AvgNsel(s.d.)}} \\
		\hline
		PIPE           & 0.310(0.336)  & 0.589(0.262) & 0.073(0.072) & 1.060(0.754) \\
		waterfall      & 0.368(0.409) & 0.705(0.166) & 0.272(0.202) & 2.037(1.391) \\
		BOIN combo     & 0.107(0.133) & 0.681(0.177) & 0.148(0.131) & 0.750(0.344) \\
		Keyboard combo & 0.103(0.131) & 0.682(0.178) & 0.151(0.132) & 0.751(0.344) \\
 
		   Ci3+3       &  0.117(0.144)   &   0.689(0.182)  &   0.124(0.109)   &   0.739(0.354)  \\



	
		\hline
	\end{tabular}
	
	\vskip 0.2in 

	\centering
	Average patient allocation in 100 scenarios.
	\begin{tabular}{c|ccc|c}
		\hline
		\textbf{Design} & \textbf{\#UA(s.d.)} & \textbf{\#CA(s.d.)} & \textbf{\#OA(s.d.)} & \textbf{Total(s.d.)} \\
		\hline
		PIPE            & 27.660(31.904)  & 29.236(20.998) & 21.819(17.186) & 78.715(24.791) \\
		waterfall       & 25.699(31.900)   & 17.968(12.133) &  12.209(8.941)  & 55.877(30.872) \\
		BOIN combo      & 19.276(19.757) & 37.936(22.326) & 20.545(15.429) & 77.757(25.584) \\
		Keyboard combo  & 19.043(19.544) & 37.726(22.184) & 21.062(15.476) & 77.832(25.564) \\

		 Ci3+3       &   17.426(17.901)   &   37.611(22.22)  &  22.939(16.17)  &  77.977(25.497)   \\



		
		\hline
	\end{tabular}

	\label{tab:sim2_sel_and_alloc}
\end{table}

	

One interesting observation is that the actual total sample size of the waterfall design is much smaller than the maximum sample size we set even if ``n.earlystop" equals to the maximum sample size in the function ``get.oc.combo". 
We give an explanation for this phenomenon in  Appendix E.

 Lastly, we investigate the performance of the Ci3+3 design with three additional modifications as part of sensitivity analyses: (1) skip stage I and start the trial from stage II , (2) use EP $P_1$ for stage I escalation, and (3) use $P_2$ for stage I escalation. The results are shown in Appendix F. The Ci3+3 design shows remarkable robustness in all three cases.

\section{Discussion}

We have proposed Ci3+3, a rule-based design for combination dose-finding trials with two agents. We show through simulation studies that the operating characteristics of Ci3+3 are comparable to existing designs in a variety of scenarios. The algorithms in the two stages of Ci3+3 are relatively simple to follow. In the first stage, once an EP is selected, dose-escalation decisions only require comparing the observed toxicity rate and the EI at the current DC. In the second stage, realizing that there are multiple DCs available for each up-and-down decisions D, E, and S, and therefore, the Ci3+3 design utilizes a utility function that prefers the DCs    with high posterior probability belonging to EI.    Statistical inference is limited to the use of a working beta/binomial model, and the entire design can be implemented with a set of rules.
\yy Implementation of Ci3+3 for practical trials is simple in the first stage, where the up-and-down decisions may be carried out using a decision table based on the preselected EP and i3+3. The second stage is not simpler than model-based designs, however, since it also requires computation of posterior probabilities based on a Bayesian model. \jj

Our attempt to simplify the design of combination dose-finding trials is motivated by the recent development of simple and interval-based dose-finding designs for single-agent dose finding. The use of EI and up-and-down decisions serves the build blocks of the Ci3+3 design, just as they did for i3+3, mTPI-2, and keyboard designs. However, due to the complex nature of two-dimensional dose finding, the Ci3+3 design cannot be reduced to a decision table as shown in the single-agent designs. For example, when 2 out of 3 patients experience DLT at a DC, a decision D is needed but such a decision does not provide the exact dose level since multiple lower DCs may be available.

Future directions include extending Ci3+3 for trials with more than two agents, with joint outcomes of efficacy and toxiciy, and delayed outcomes. Also, another extension could consider allowing multiple EPs to enroll patients simultaneously in Stage I, which would potentially result in time saving and higher efficiency of dose finding.

\clearpage
\newpage

\bibliographystyle{apalike}
\bibliography{Ci3+3}

\begin{thebibliography}{}

\bibitem[Braun and Jia, 2013]{braun2013generalized}
Braun, T.~M. and Jia, N. (2013).
\newblock A generalized continual reassessment method for two-agent phase i
  trials.
\newblock {\em Statistics in biopharmaceutical research}, 5(2):105--115.

\bibitem[Bril et~al., 1984]{biviso}
Bril, G., Dykstra, R., Pillers, C., and Robertson, T. (1984).
\newblock Algorithm as 206: Isotonic regression in two independent variables.
\newblock {\em Journal of the Royal Statistical Society. Series C (Applied
  Statistics)}, 33(3):352--357.

\bibitem[Cheung, 2011]{cheung2011dose}
Cheung, Y.~K. (2011).
\newblock {\em Dose finding by the continual reassessment method}.
\newblock CRC Press.

\bibitem[Conaway et~al., 2004]{conaway2004designs}
Conaway, M.~R., Dunbar, S., and Peddada, S.~D. (2004).
\newblock Designs for single-or multiple-agent phase i trials.
\newblock {\em Biometrics}, 60(3):661--669.

\bibitem[Korn and Simon, 1993]{korn1993using}
Korn, E.~L. and Simon, R. (1993).
\newblock Using the tolerable-dose diagram in the design of phase i combination
  chemotherapy trials.
\newblock {\em Journal of Clinical Oncology}, 11(4):794--801.

\bibitem[Kramar et~al., 1999]{kramar1999continual}
Kramar, A., Lebecq, A., and Candalh, E. (1999).
\newblock Continual reassessment methods in phase i trials of the combination
  of two drugs in oncology.
\newblock {\em Statistics in medicine}, 18(14):1849--1864.

\bibitem[Lin and Yin, 2017]{lin2017bayesian}
Lin, R. and Yin, G. (2017).
\newblock Bayesian optimal interval design for dose finding in drug-combination
  trials.
\newblock {\em Statistical methods in medical research}, 26(5):2155--2167.

\bibitem[Liu and Yuan, 2015]{2015Bayesian}
Liu, S. and Yuan, Y. (2015).
\newblock {Bayesian optimal interval designs for phase I clinical trials}.
\newblock {\em Journal of the Royal Statal Society: Series C (Applied Stats)},
  64.

\bibitem[Mander and Sweeting, 2015]{mander2015product}
Mander, A.~P. and Sweeting, M.~J. (2015).
\newblock A product of independent beta probabilities dose escalation design
  for dual-agent phase i trials.
\newblock {\em Statistics in medicine}, 34(8):1261--1276.

\bibitem[Neuenschwander et~al., 2015]{neuenschwander2015bayesian}
Neuenschwander, B., Matano, A., Tang, Z., Roychoudhury, S., Wandel, S., and
  Bailey, S. (2015).
\newblock A bayesian industry approach to phase i combination trials in
  oncology.
\newblock {\em Statistical methods in drug combination studies}, 2015:95--135.

\bibitem[Pan et~al., 2017]{pan2017statistical}
Pan, H., Lin, R., and Yuan, Y. (2017).
\newblock Statistical properties of the keyboard design with extension to
  drug-combination trials.

\bibitem[Tighiouart et~al., 2017]{Tighiouart2017ABA}
Tighiouart, M., Li, Q., and Rogatko, A. (2017).
\newblock A bayesian adaptive design for estimating the maximum tolerated dose
  curve using drug combinations in cancer phase i clinical trials.
\newblock {\em Statistics in medicine}, 36 2:280--290.

\bibitem[Tighiouart et~al., 2014]{Tighiouart2014DoseFW}
Tighiouart, M., Piantadosi, S., and Rogatko, A. (2014).
\newblock Dose finding with drug combinations in cancer phase i clinical trials
  using conditional escalation with overdose control.
\newblock {\em Statistics in medicine}, 33 22:3815--29.

\bibitem[Wages et~al., 2011]{wages2011continual}
Wages, N.~A., Conaway, M.~R., and O'Quigley, J. (2011).
\newblock Continual reassessment method for partial ordering.
\newblock {\em Biometrics}, 67(4):1555--1563.

\bibitem[Wages et~al., 2014]{2014Phase}
Wages, N.~A., O"Quigley, J., and Conaway, M.~R. (2014).
\newblock Phase i design for completely or partially ordered treatment
  schedules.
\newblock {\em Statistics in Medicine}, 33(4):569--579.

\bibitem[Yan et~al., 2017]{yan2017keyboard}
Yan, F., Mandrekar, S.~J., and Yuan, Y. (2017).
\newblock {Keyboard: a novel {B}ayesian toxicity probability interval design
  for phase {I} clinical trials}.
\newblock {\em Clinical Cancer Research}, 23(15):3994--4003.

\bibitem[Zhang and Yuan, 2016]{zhang2016practical}
Zhang, L. and Yuan, Y. (2016).
\newblock A practical bayesian design to identify the maximum tolerated dose
  contour for drug combination trials.
\newblock {\em Statistics in medicine}, 35(27):4924--4936.

\end{thebibliography}

\newpage
\section*{ Appendix A}\label{app:A}

\renewcommand{\thefigure}{A.\arabic{figure}}
\renewcommand{\thetable}{A.\arabic{table}}
\setcounter{figure}{0}
\setcounter{table}{0}

\begin{table}[H]
	\begin{center}
	  \caption{The i3+3 design and its decision rules. Here, the term ``below the EI'' means that ``less than the lower bound ($p_T-\epsilon_1$) of the EI''; ``above the EI'' means that ``greater than the upper bound ($p_T+\epsilon_2$) of the EI'' and ``inside the EI'' means inbetween the two bounds.    } \label{tab:i3+3} 
	  \begin{tabular}{p{7cm} {c} r r}
		\hline
		\multicolumn{3}{c}{Current dose: $d$; Patients treated: $n_d$; Patients with DLTs: $y_d$} \\
		\hline
		Condition & Decision& Dose for the next cohort\\
		\hline
		$\frac{y_d}{n_d}$ below the EI	& Escalation ($E$) & $d+1$\\
		$\frac{y_d}{n_d}$ inside the EI	& Stay ($S$) & $d$ \\
		$\frac{y_d}{n_d}$ above the EI and $\frac{y_d-1}{n_d}$ below the EI	& Stay ($S$)  & $d$\\
		$\frac{y_d}{n_d}$ above the EI and $\frac{y_d-1}{n_d}$ inside the EI	& De-escalation ($D$)  & $d-1$\\
		$\frac{y_d}{n_d}$ above the EI and $\frac{y_d-1}{n_d}$ above the EI	& De-escalation ($D$)  & $d-1$\\
		\hline
	  \end{tabular}
	\end{center}
  \end{table}
  
  \begin{figure}[H]
	  \centering \includegraphics[scale=0.43]{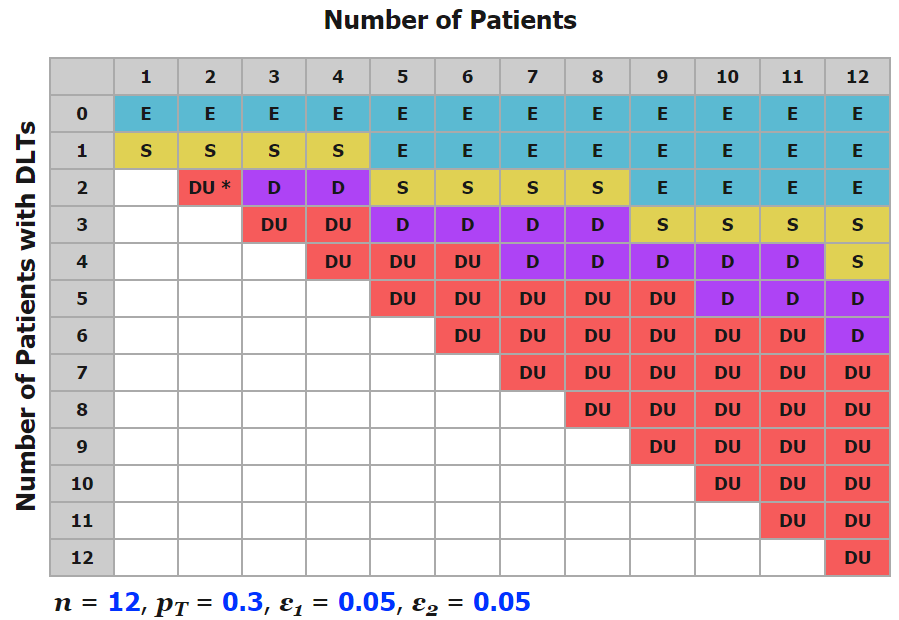} 
	  \caption{An example of the i3+3 dose-escalation decision table    for 12 patients.    The target toxicity probability $p_T=0.3$, and the equivalence interval (EI) is $(0.25,0.35)$.  Each column represents $n_{d}$ number of subjects treated at the current dose $d$  and each row represents      $y_d$  number of subjects with DLTs at the current    dose.     Each cell in the table provides the dose-escalation decision based on the readouts from the corresponding row ($y_{ij}$) and column ($n_{ij}$), where \textbf{E} denotes escalate to the next higher dose, \textbf{S} stay at the same dose, \textbf{D} de-escalate to the previous lower dose, and \textbf{DU} de-escalate to the previous lower dose, and the current    and higher doses are excluded and    will never be used again in the trial. } \label{fig:i3+3-dtab}
  \end{figure}

  \section*{ Appendix B}
  \renewcommand{\thefigure}{B.\arabic{figure}}
  \renewcommand{\thetable}{B.\arabic{table}}
  \setcounter{figure}{0}
  \setcounter{table}{0}

    We consider modify the orginal Ci3+3 algorithm by adding the special case when  $\cA_{ij}=S$ and $n_{ij} >= 12$.  We simulate the modified Ci3+3 using scenarios in Simulation Study 2.  From Table \ref{tab:sim2_sel_and_alloc_without12}, we can see that in general, compared to the original Ci3+3, the OA and POS of the modified Ci3+3 is higher. This is because the modified Ci3+3 explores the untested DCs in the candidate set even if the current DC has demonstrated a high probability of being the MTDC, thus resulting in higher OA and POS values. The advantage of the modified Ci3+3 is that it can explore more DCs. We can see that the average number of DCs used in the modified Ci3+3 design is slightly higher than that in the original Ci3+3 design.

  
	  

  \begin{table}[!htbp]
	  \centering
	  \caption{Simulation results for Simulation Study 2.}
	  Average MTDC selection in 100 scenarios.
	  \begin{tabular}{c|ccc|c}
		  \hline
		  \textbf{Design} & \textbf{PUS(s.d.)} & \textbf{PCS(s.d.)} & \textbf{POS(s.d.)} & \multicolumn{1}{c}{\textbf{AvgNsel(s.d.)}} \\
		  \hline
   
		  Original Ci3+3      & 0.117(0.144)  & 0.689(0.182) & 0.124(0.109) & 0.739(0.354) \\
		  Modified Ci3+3      & 0.111(0.141)  & 0.680(0.187) & 0.140(0.121) & 0.740(0.354) \\

		  \hline
	  \end{tabular}
	  
	  \vskip 0.2in 
  
	  \centering
	  Average patient allocation in 100 scenarios.
	  \begin{tabular}{c|ccc|c}
		  \hline
		  \textbf{Design} & \textbf{\#UA(s.d.)} & \textbf{\#CA(s.d.)} & \textbf{\#OA(s.d.)} & \textbf{Total(s.d.)} \\
		  \hline
   
		  Original Ci3+3      & 17.426(17.901)  & 37.611(22.22) & 22.939(16.17) & 77.977(25.497)  \\
		  Modified Ci3+3      & 16.947(17.465)  & 37.302(22.21) & 23.809(16.341) & 78.058(25.514)  \\
		     
		  \hline
	  \end{tabular}

	  \vskip 0.2in 
  
	  \centering
	  Average number of DCs used in 100 scenarios.

	  \begin{tabular}{c|ccc|c}
		  \hline
		  \textbf{Design} & \textbf{\#DCs used(s.d.)}   \\
		  \hline
   
		  Original Ci3+3      & 4.421(2.207)      \\
		  Modified Ci3+3      & 5.256(2.717)    \\
		     
		  \hline
	  \end{tabular}
  
	  \label{tab:sim2_sel_and_alloc_without12}
  \end{table}

\section*{ Appendix C}

\renewcommand{\thefigure}{C.\arabic{figure}}
\renewcommand{\thetable}{C.\arabic{table}}
\setcounter{figure}{0}
\setcounter{table}{0}

Below are the summary of the 100 scenarios in Simulation Study 2.
\begin{table}[H]
	\centering
	\caption{Possible   toxicity probabilities for a single agent of four doses.  }
	\begin{tabular}{c|rrrr}
		\hline
		\multicolumn{1}{c|}{\multirow{2}[3]{*}{\textbf{Scenario}}} & \multicolumn{4}{c}{\textbf{Dose Level}} \\
		\cline{2-5}          
		& 1     & 2     & 3     & 4 \\
		\hline
		1     & 0.15  & 0.3   & 0.45  & 0.6 \\
		2     & 0.1   & 0.2   & 0.3   & 0.4 \\
		3     & 0.08  & 0.16  & 0.24  & 0.44 \\
		4     & 0.06  & 0.12  & 0.18  & 0.24 \\
		5     & 0.26  & 0.38  & 0.5   & 0.62 \\
		\hline
	\end{tabular}
	\label{tab:single-agent}
\end{table}

\begin{table}[H]
	\centering
	\caption{Possible values of interaction coefficient. Protective means that the drug combination produces a toxic effect less than that if the drugs act independently in the body; Synergistic means the drug combination produces a toxic effect greater than that if the drugs act independently in the body.}
	\begin{tabular}{c|c}
		\hline
		$\bm{\eta}$  & \multicolumn{1}{c}{\textbf{Meaning}}  \\
		\hline
		-2    & Strongly protective\\
		-0.2  &  Strongly protective\\
		0.2   & Weakly synergistic\\
		0.7     & Strongly synergistic\\
		\hline
	\end{tabular}
	\label{tab:inter}
\end{table}

\begin{table}[H]
	\centering
	\caption{Summary of 100 scenarios in Simulation 2. With $p_T = 0.3$ and $\epsilon_1 = \epsilon_2 = 0.05$, the safe scenarios mean that all DCs are safe and tolerable,   that is, $p_{ij}<0.25$, for all $d_{ij}$. The toxic scenarios mean that all DCs are overly toxic, that is, $p_{ij}>0.35$, for all DC $d_{ij}$. The other scenarios possess different numbers of MTDCs.   }
	\begin{tabular}{cc|c}
		\hline
		\multicolumn{2}{c|}{\textbf{Scenario Category}} & \multicolumn{1}{c}{\textbf{Number}} \\
		\hline
		\cline{1-3}   
		& All DCs are safe  & 13 \\
		& With 1 MTDC     & 18 \\
		& With 2 MTDCs    & 24 \\
		& With 3 MTDCs    & 5 \\
		& With more than 3 MTDCs    & 18 \\
		& All DCs are toxic & 22 \\
		\cline{1-3}  
		\hline
		\multicolumn{2}{c|}{\textbf{Total}} & 100 \\
		\hline
	\end{tabular}
	\label{tab:two-agent}
\end{table}

\section*{ Appendix D}
\renewcommand{\thefigure}{D.\arabic{figure}}
\renewcommand{\thetable}{D.\arabic{table}}
\setcounter{figure}{0}
\setcounter{table}{0}

\begin{figure}[H]
	\centering \includegraphics[width=\textwidth]{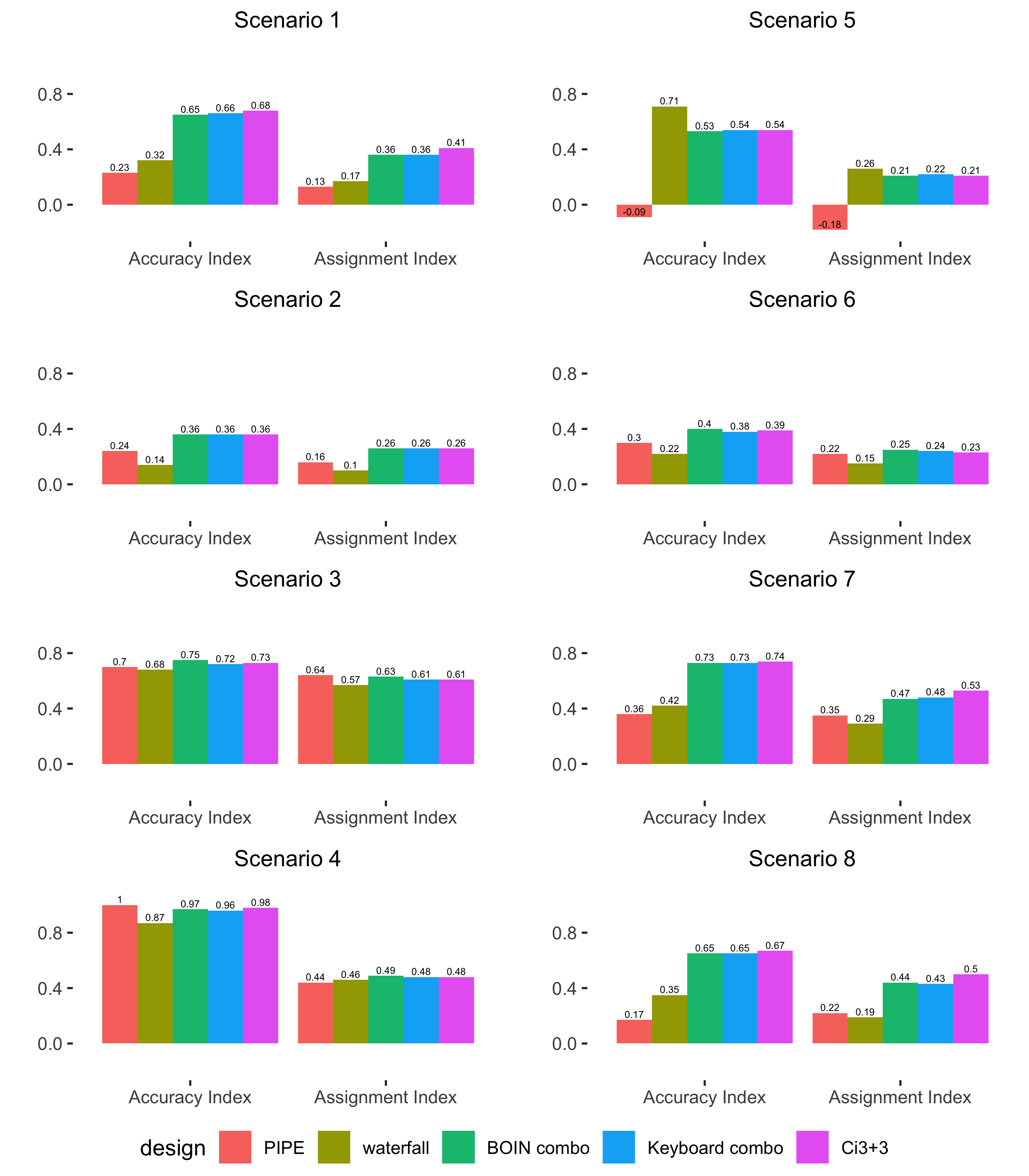}
	\caption{The Accuracy Index and Assignment Index in Simulation Study 1.}
	\label{fig:accuracy_assignment_index}
\end{figure}

Figure \ref{fig:accuracy_assignment_index} and Table \ref
{tab:accuracy_assignment_index} show the Accuracy Index and 
Assignment Index for the PIPE, waterfall, BOIN combo, Keyboard combo and Ci3+3 designs in Simulation Studies 1 $\&$ 2, respectively. When calculating the Accuracy Index, we randomly select one DC among the selected MTDC if PIPE and waterfall select more than one DC as the MTDC. In general, the Accuracy Index and Assignment Index of Ci3+3 are higher than that of PIPE and waterfall and are comparable to that of BOIN combo and Keyboard combo.

\begin{table}[!htbp]
	\centering
	\caption{The Accuracy Index and Assignment Index in Simulation Study 2.}
	Average Accuracy Index and Assignment Index in 100 scenarios.
	\begin{tabular}{c|ccc|c}
		\hline
		\textbf{Design} & \textbf{Accuracy Index(s.d.)} & \textbf{Assignment Index(s.d.)}\\
		\hline
		PIPE           & 0.688(0.286) & 0.498(0.225)  \\
		waterfall      & 0.642(0.258) & 0.483(0.233)   \\
		BOIN combo     & 0.750(0.187) & 0.558(0.196)   \\
		Keyboard combo & 0.750(0.186) & 0.555(0.194) \\
		 Ci3+3   &  0.754(0.191)    &  0.550(0.188)   \\

		\hline
	\end{tabular}

	\label{tab:accuracy_assignment_index}
\end{table}

\section*{ Appendix E}
\renewcommand{\thefigure}{E.\arabic{figure}}
\renewcommand{\thetable}{E.\arabic{table}}
\setcounter{figure}{0}
\setcounter{table}{0}

\begin{figure}[!htbp]
	\centering \includegraphics[scale=0.4]{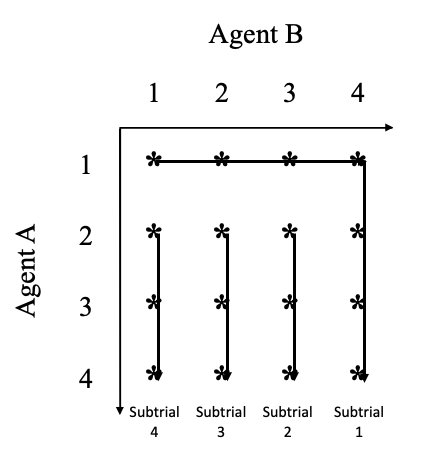}
	\caption{An example of subtrials in the waterfall design.}
	\label{fig:waterfall_subtrial}
\end{figure}

  We find that for the waterfall design, in some scenarios, sample size may be small because certain subtrials are omitted without any patients enrolled.     For example, with 4 dose levels of both agents A and B,   waterfall divides the trial into four subtrials with recommending sample sizes, as shown in Figure \ref{fig:waterfall_subtrial}.   In scenario 3 of Simulation Study 1, the true toxicity probability 0.25 of DC $d_{12}$ is close to $p_T$ and it is one of the MTDCs. At this time, if Subtrial 1 identified $d_{12}$ as the candidate MTDC, Subtrial 4 will be conducted next and Subtrials 2 and 3 will be omitted and never conducted.

\section*{ Appendix F}
\renewcommand{\thefigure}{F.\arabic{figure}}
\renewcommand{\thetable}{F.\arabic{table}}
\setcounter{figure}{0}
\setcounter{table}{0}

Table \ref{tab:sim2_sel_and_alloc_ci3_different_settings} shows the performance of the Ci3+3 design with different settings in Simulation Study 2. Ci3+3-stageII means the design starts a trial from Stage II directly by skipping Stage I; Ci3+3-P1 means the Ci3+3 design $P_1$ as the first-stage escalation path; and Ci3+3-P2 means $P_2$. The performance of the four designs are very similar.

\begin{table}[!htbp]
	\centering
	\caption{Simulation results of Ci3+3 with different settings for Simulation Study 2. Ci3+3 is exactly the same as the one in Simulation Study 2. All settings of the other three designs are the same as Ci3+3, except that Ci3+3-stageII is the Ci3+3 design skipping stage I and starting from stage II, Ci3+3-P1 is the Ci3+3 design using $P_1$ for stage I excalation, and Ci3+3-P2 is the Ci3+3 design using $P_2$ for stage I excalation.}
	Average MTDC selection in 100 scenarios.
	\begin{tabular}{c|ccc|c}
		\hline
		\textbf{Design} & \textbf{PUS(s.d.)} & \textbf{PCS(s.d.)} & \textbf{POS(s.d.)} & \multicolumn{1}{c}{\textbf{AvgNsel(s.d.)}} \\
		\hline
 
		   Ci3+3       &  0.117(0.144)   &   0.689(0.182)  &   0.124(0.109)   &   0.739(0.354)  \\

		   Ci3+3-stageII      &  0.122(0.150)   &   0.684(0.183)  &   0.123(0.109)   &   0.737(0.354)  \\

		   Ci3+3-P1       &  0.122(0.150)   &   0.685(0.182)  &   0.123(0.109)   &   0.738(0.353)  \\

		   Ci3+3-P2       &  0.123(0.152)   &   0.684(0.183)  &   0.123(0.109)   &   0.738(0.354)  \\
		
		\hline
	\end{tabular}
	
	\vskip 0.2in 

	\centering
	Average patient allocation in 100 scenarios.
	\begin{tabular}{c|ccc|c}
		\hline
		\textbf{Design} & \textbf{\#UA(s.d.)} & \textbf{\#CA(s.d.)} & \textbf{\#OA(s.d.)} & \textbf{Total(s.d.)} \\
		\hline

		 Ci3+3       &   17.426(17.901)   &   37.611(22.22)  &  22.939(16.17)  &  77.977(25.497)   \\

		 Ci3+3-stageII       &   17.531(18.195)   &   37.588(22.122)  &  22.813(16.123)  &  77.932(25.500)   \\

		 Ci3+3-P1       &   17.912(18.678)   &  37.267(22.031)  &  22.763(16.157)  &  77.942(25.476)   \\

		 Ci3+3-P2       &   17.869(18.647)   &   37.251(22.003)  &  22.829(16.171)  &  77.949(25.493)   \\

		\hline
	\end{tabular}

	\label{tab:sim2_sel_and_alloc_ci3_different_settings}
\end{table}

\end{document}